# New room temperature multiferroics on the base of single-phase nanostructured perovskites

Maya D. Glinchuk[1], Eugene A. Eliseev[1] and Anna N. Morozovska[1,2*]

[1] Institute for Problems of Materials Science, NAS of Ukraine,
Krjijanovskogo 3, 03142 Kiev, Ukraine

[2] Institute of Physics, NAS of Ukraine, 46, pr. Nauki, 03028 Kiev, Ukraine

**Abstract**

The theoretical description of the nanostructured $Pb(Fe_{1/2}Ta_{1/2})_x(Zr_{1/2}Ti_{1/2})_{1-x}O_3$ (**PFTx-PZT(1-x)**) and $Pb(Fe_{1/2}Nb_{1/2})_x(Zr_{1/2}Ti_{1/2})_{1-x}O_3$ (**PFNx-PZT(1-x)**) intriguing ferromagnetic, ferroelectric and magnetoelectric properties at temperatures higher than 100 K are absent to date. The goal of this work is to propose the theoretical description of the physical nature and the mechanisms of the aforementioned properties, including room temperature ferromagnetism, phase diagram dependence on the composition x with a special attention to the multiferroic properties at room temperature, including anomalous large value of magnetoelectric coefficient. The comparison of the developed theory with experiments establishing the boundaries between paraelectric, ferroelectric, paramagnetic, antiferromagnetic, ferromagnetic and magnetoelectric phases, as well as the characteristic features of ferroelectric domain switching by magnetic field are performed and discussed. The experimentally established absence of ferromagnetic phase in PFN, PFT and in the solid solution of PFN with $PbTiO_3$ (**PFNx-PT(1-x)**) was explained in the framework of the proposed theory.

## 1. Introduction

The search of room temperature magnetoelectric multiferroics is known to be a hot topic for researchers and engineers working in the field of novel functional devices fabrication [1, 2, 3, 4, 5]. For the majority of these devices operation at room temperature and significant magnetoelectric coupling are especially vital. Until recently such characteristics were demonstrated on multiferroic heterostructures [6, 7, 8, 9]. The discovery of single-phase room temperature magnetoelectrics on the basic of solid solutions of ferroelectric antiferromagnets $Pb(Fe_{1/2}Ta_{1/2})O_3$ (**PFT**) and $Pb(Fe_{1/2}Nb_{1/2})O_3$ (**PFN**) with $Pb(Zr_{1/2}Ti_{1/2})O_3$ (**PZT**) seems to be very important [6, 8, 9]. Since we will pay attention to the solid solutions $Pb(Fe_{1/2}Ta_{1/2})_x(Zr_{1/2}Ti_{1/2})_{1-x}O_3$ (**PFTx-PZT(1-x)**) and $Pb(Fe_{1/2}Nb_{1/2})_x(Zr_{1/2}Ti_{1/2})_{1-x}O_3$ (**PFNx-PZT(1-x)**) in this paper, let us discuss briefly their properties at $T \geq 100\,K$.

PFN is an antiferromagnet with G-type spin ordering below at $T<T_{Neel}$, where $T_{Neel}$=143-170 K [10, 11, 12,]. Also it is conventional ferroelectric at temperatures $T<T_{Curie}$. Ferroelectric phase transition (as reported in different works) appears in the range $T_{Curie}$=379 − 393K [11, 12, 13, 14]. PFN has biquadratic magnetoelectric (ME) coupling constant $2.2\times10^{-22}$ sm/(VA) at 140K [15]. PFT is an antiferromagnet with

---
* Corresponding author: anna.n.morozovska@gmail.com

Neel temperature $T_{Neel}$= 133-180 K [16, 17, 18] and ferroelectric with the phase transition at $T_{Curie} \approx 250$ K [19], at that the value is slightly dependent on the external field frequency. PFT has biquadrartic ME coupling constant of the same order as PFN. PZT is nonmagnetic and conventional ferroelectric with Curie transition temperature varied in the range 666-690 K depending on the sample preparation [20].

PFTx-PZT(1-x) and PFNx-PZT(1-x) was studied at composition x=0.1 – 0.4 [6, 8, 9]. At x=0.1 ferromagnetism is faint, while at x=0.2 – 0.4 PFTx-PZT(1-x) exhibits square saturated magnetic hysteresis loops with magnetization 0.1 emu/g at 295 K and pronounced saturated ferroelectric hysteresis with saturation polarization 25 $\mu C/cm^2$, which actually increases to 40 $\mu C/cm^2$ in the high temperature tetragonal phase, representing an exciting new room temperature oxide multiferroic [8, 9]. Giant effective ME coefficient of PFTx-PZT(1-x) was reported as $1.3 \times 10^{-7}$ s/m for x=0.3 and 0.4 [8], however it appeared to be a *nonlinear* effect. Meanwhile Ref. [9] reports about much smaller value of the *linear* ME coefficient, $1.3 \times 10^{-11}$ s/m. The reason of the strong discrepancy between the ME coefficients was not explained and stays unclear.

PFNx-PZT(1-x) demonstrates magnetization loop vs. applied magnetic field at room temperature for x between 0.1 and 0.4, however an improvement in ferromagnetic properties was observed for x=0.2 and x=0.3, while a notable deterioration of the these properties was observed for x=0.1 and 0.4. [6, 9]. Saturated and low loss ferroelectric hysteresis curves with a remanent polarization of about 20-30 $\mu C/cm^2$ was observed in Refs.[6, 9, 21].

Note that Scott [5] stressed that the magnetoelectric switching of single-phase nanocrystals in PFTx-PZT(1-x) was reported by Evans et al [8]. Actually PFTx-PZT(1-x) samples studied there have a lamellar structure (of 150 nm width) with pronounced nanodomains of about 10-50 nm average diameter. These are known to be slightly Fe-rich nanoregions, but not a different phase in PFTx-PZT(1-x) [5]. Sanchez et al [9] stated that Fe spin clustering plays a key role in the room-temperature magnetoelectricity of these materials. In such case it is useful to have evidence that they may be considered as single-phase crystals rather than nanocomposites. Moreover, Sanchez et al [6] revealed that the local phonon mode $A_{1g}$ corresponds to ordered nanodomains in PFNx-PZT(1-x) and it is attributed to the vibration of oxygen ion in the oxygen octahedra. These facts speaks in favour that the presence of nanostructure can play an important role in the description of solid solutions PFTx-PZT(1-x) and PFNx-PZT(1-x) physical properties, including primary the room temperature ferromagnetism. So that it seems prospective to consider nanostructured PFNx-PZT(1-x) and PFTx-PZT(1-x), where the term "nanostructured" suppose either Fe-richer nanoregions, nanocrystals or nanosized lamellas, or by extension, artificially created nanograined ceramics. In what follows we will call any of them nanoregions.

To the best of our knowledge there are no published papers devoted to the theoretical description of the nanostructured PFNx-PZT(1-x) and PFTx-PZT(1-x) intriguing ferromagnetic, ferroelectric and magnetoelectric properties at temperatures $T \geq 100$ K. Therefore the main goal of this work is to propose the theoretical description of the physical nature and mechanisms of the aforementioned properties, including room temperature ferromagnetism, phase diagram dependence on the composition x with a special attention to the multiferroic properties at room temperature. The comparison of the developed theory with experiments establishing the boundaries between paraelectric (**PE**), paramagnetic (**PM**), antiferromagnetic (**AFM**), ferroelectric (**FE**), ferromagnetic (**FM**) and magnetoelectric (**ME**) phases, as well as characteristic features of the ferroelectric domain switching by magnetic field are performed and discussed.

## 2. Landau-Ginzburg model

The polarization and structural parts of the 2-4-power Landau-potential homogeneous bulk density is the sum of polarization ($g_P$), antimagnetization ($g_L$), magnetization ($g_M$), elastic ($g_{el}$) and magnetoelectric ($g_{ME}$) parts [22]:

$$G_{PM} = g_P + g_L + g_M + g_{el} + g_{ME} \qquad (1)$$

The densities $g_P = \frac{\alpha_P}{2}P_i^2 + \frac{\beta_{Pij}}{4}P_i^2 P_j^2 + q_{ijkl}^{(e)} u_{ij} P_k P_l$, $g_L = \frac{\alpha_L}{2}L_i^2 + \frac{\beta_{Lij}}{4}L_i^2 L_j^2 + q_{ijkl}^{(m)} u_{ij} L_k L_l$,

$g_M = \frac{\alpha_M}{2}M_i^2 + \frac{\beta_{Mij}}{4}M_i^2 M_j^2 + q_{ijkl}^{(l)} u_{ij} M_k M_l$, $g_{el} = \frac{c_{ijkl}}{2} u_{ij} u_{kl} + \frac{A_{ijklmn}}{2} u_{ij} u_{kl} P_m P_n + \frac{B_{ijklmn}}{2} u_{ij} u_{kl} L_m L_n + \frac{C_{ijklmn}}{2} u_{ij} u_{kl} M_m M_n$.

**P** is the polarization, $L_i = (M_{ai} - M_{bi})/2$ is the components of antimagnetization vector of two equivalent sub-lattices $a$ and $b$, and $M_i = (M_{ai} + M_{bi})/2$ is the magnetization vector components; $u_{ij}$ is elastic strain tensor; $q_{ijkl}^{(e)}$, $q_{ijkl}^{(l)}$ and $q_{ijkl}^{(m)}$ the bulk electro- and magnetostriction coefficients, $c_{ijkl}$ are elastic stiffness.

The linear and quadratic magnetoelectric (ME) energy is

$$g_{ME} = \mu_{ij} M_i P_j + \frac{\eta_{ijkl}^{FM}}{2} M_i M_j P_k P_l + \frac{\eta_{ijkl}^{AFM}}{2} L_i L_j P_k P_l, \qquad (2a)$$

$\mu_{ij}$ is the bilinear ME coupling term, $\eta_{ijkl}^{FM}$ and $\eta_{ijkl}^{AFM}$ are the components of the biquadratic ME coupling term. The last two terms in Eq.(2) can be obtained via electro- and magneto-striction terms as it was shown in [23]:

$$\eta_{ijkl}^{FM}(R) = -\eta_{ijkl} + \left(q_{ijmn}^{(e)} s_{mnsp} q_{spkl}^{(m)} - \left(\widetilde{A}_{ijsp} g_{ksn}^{(m)} g_{lpn}^{(m)} + \widetilde{B}_{ijsp} g_{ksn}^{(e)} g_{lpn}^{(e)}\right)\right)\left(1 + \frac{R_{\mu 1}}{R} + \left(\frac{R_{\mu 2}}{R}\right)^2\right), \qquad (2b)$$

$$\eta_{ijkl}^{AFM}(R) = \eta_{ijkl} + \left(q_{ijmn}^{(e)} s_{mnsp} q_{spkl}^{(l)} - \left(\widetilde{C}_{ijsp} g_{ksn}^{(m)} g_{lpn}^{(m)} + \widetilde{D}_{ijsp} g_{ksn}^{(e)} g_{lpn}^{(e)}\right)\right)\left(1 + \frac{R_{\mu 1}}{R} + \left(\frac{R_{\mu 2}}{R}\right)^2\right). \qquad (2c)$$

Here $\eta_{ijkl}$ is the "bare" ME coupling tensor, $s_{mnsq}$ are elastic compliances, $g_{ijk}^{(e)}$ and $g_{ijk}^{(m)}$ are tensors of piezoelectric and piezomagnetic effects respectively. One can see from the Eqs.(2b,c) that $\left|\eta_{ijkl}^{AFM}\right| \neq \left|\eta_{ijkl}^{FM}\right|$ due to the difference in electrostriction and (anti)magnetostriction coefficients $q_{ijkl}^{(m)} \neq q_{ijkl}^{(l)}$ and different higher coupling constants $\tilde{A}_{ijkl} \neq \tilde{C}_{ijkl}$ and $\tilde{B}_{ijkl} \neq \tilde{D}_{ijkl}$. Moreover, since the striction, piezoelectric and piezomagnetic tensors strongly depend on the composition x, the ME coupling coefficients $\eta_{ijkl}^{FM}$ and $\eta_{ijkl}^{AFM}$ can vary strongly for PFN and PFT.

Hereinafter $R$ is the average size of the nanoregion. The characteristic radii $R_{\mu 1}$ and $R_{\mu 2}$ originate from the intrinsic surface stress effect and its value is proportional to the product of the surface tension coefficient $\mu$, electrostriction $Q_{ij}$, magnetostrictriction $Z_{ij}$, nonlinear electro- and magnetostriction coefficients $A_{ij}$ and $B_{ij}$ correspondingly in Voight notations. Namely $R_{\mu 1} \propto \mu \left( \frac{Q_{ij} B_{ij} + Z_{ij} A_{ij}}{Q_{ij} Z_{ij}} \right)$ and $R_{\mu 2} \propto \mu \sqrt{\frac{A_{ij} B_{ij}}{Q_{ij} Z_{ij}}}$ (see details in Ref.[23]). The characteristic radii value depends on the nanoregion shape. In particular case of nanoregions observed in [8], it can be modeled by a wire of radius $R$, radii $R_{\mu 1} = 2\mu \frac{s_{12}}{s_{11}} \left( \frac{Q_{11} B_{33} + Z_{11} A_{33}}{Q_{11} Z_{11}} \right)$ and $R_{\mu 2} = 2\mu \frac{s_{12}}{s_{11}} \sqrt{\frac{A_{33} B_{33}}{Q_{11} Z_{11}}}$, where $s_{ij}$ are elastic compliances in Voight notations (see Eq.(9c) in Ref.[23]). It was shown that $R_{\mu 1}$ and $R_{\mu 2}$ values can reach several hundreds of nm and so the contribution $(R_{\mu 1}/R) + (R_{\mu 2}/R)^2$ in Eqs.(2b,c) may increase the ME coupling coefficient in 10 – $10^3$ times for the average size $R \propto 20 - 2$ nm (see figure 3 in Ref.[23]).

The coefficient $\alpha_P$ linearly depends on temperature, i.e. $\alpha_P = \alpha_{TY}\left(T - T_Y^C\left(1 - \frac{R_{PY}}{R}\right)\right)$ [23], where Y="N" or "T" and $T_Y^C$ is the ferroelectric transition temperature of homogeneous bulk. In the work [22] the form of coefficients $\alpha_L$ and $\alpha_M$ were obtained for any antiferromagnet with two sublattices $a$ and $b$. It is a common knowledge (see e.g. [24]) that with account of an exchange interaction constant $c$ between the sub-lattices and the interaction constant inside sublattices $b$ two characteristic temperatures have to be introduced: Neel temperature, $T_N = \frac{(c-b)}{2\alpha_M^T}$, that defines the magnetic susceptibility behavior at $T \leq T_N$, and Curie temperature, $\theta_C = -\frac{(c+b)}{2\alpha_M^T}$, that defines the magnetic susceptibility behavior at $T > T_N$. Because of this one can rewrite the expressions for the coefficients obtained in [21, 23] for the case PFT and PFN in the form $\alpha_L = \alpha_{LT}\left(T - T_N^Y\left(1 - \frac{R_{LY}}{R}\right)\right)$ and $\alpha_M = \alpha_{MT}\left(T - \theta_C^Y\left(1 - \frac{R_{MY}}{R}\right)\right)$. Temperatures $T_N^Y$ and $\theta_C^Y$ correspond to the homogeneous bulk.

Similarly to the characteristic size $R_{\mu 1}$ that govern the size dependence of ME coefficients (2b,c), the critical sizes $R_{PY}$, $R_{LY}$ and $R_{MY}$ originate from the surface tension effect coupled with electrostriction and magnetostriction. The values of $R_{PY}$ and $R_{LY}$, $R_{MY}$ are proportional to the product of the surface tension coefficient and the electrostriction or magentostriction tensor coefficients correspondingly, namely

$$R_{PY} \propto \mu\left(\frac{Q_{ij}}{\alpha_{TY} T_Y^C}\right), \quad R_{LY} \propto \mu\left(\frac{\widetilde{Z}_{ij}}{\alpha_{LT} T_N^{PFY}}\right) \text{ and } R_{MY} \propto \mu\left(\frac{Z_{ij}}{\alpha_{MT} \theta_C^Y}\right)$$

(see details in Ref.[23]). In particular case of a nanowire the radii

$$R_{PY} = 2\mu\left(\frac{Q_{12} - Q_{11}(s_{12}/s_{11})}{\alpha_{TY} T_Y^C}\right), \quad R_{LY} \propto \mu\left(\frac{\widetilde{Z}_{12} - \widetilde{Z}_{11}(s_{12}/s_{11})}{\alpha_{LT} T_N^{PFY}}\right) \text{ and }$$

$$R_{MY} = 2\mu\left(\frac{Z_{12} - Z_{11}(s_{12}/s_{11})}{\alpha_{MT} \theta_C^Y}\right)$$

(see Eqs.(6a, 6b) in Ref.[23]). The critical sizes can be positive or negative depending on the nanoregion shape, electro- and magnetostriction tensor coefficients sign and surface stress direction (compressive or tensile). Their typical values are 1-10 nm [23].

Note, that all the quantities can depend on the composition x of the solid solutions PFNx-PZT(1-x) and PFTx-PZT(1-x).

### 3. PZT-PFT and PZT-PFN phase diagrams
#### 3.1. Analytical formalism

For the solid solutions the transition temperature from the PE to FE phase can be modeled using a linear law [20]:

$$T_{PFY-PZT}^{FE}(x,R) = x T_{PFY}(R) + (1-x) T_{PZT}(R) \qquad (3)$$

Hereinafter Y=N for PFNx-PZT(1-x) or Y=T for PFTx-PZT(1-x). The temperatures $T_{PZT} = 666 - 690$ K, $T_{PFN} = 383 - 393$ K and $T_{PFT} = 247 - 256$ K are defined for homogeneous bulk material with error margins depending on the sample preparation. For nanostructured material the temperatures become $R$-size dependent as $T_{PFY}(R) = T_Y^C\left(1 - \frac{R_{PY}}{R}\right)$ and $T_{PZT}(R) = T_{PZT}^C\left(1 - \frac{R_{PZT}}{R}\right)$.

In order to describe the x-composition dependence of the AFM-PM and FM-PM phase transition temperatures we will use the approach [25] based on the percolation theory [26]. We assume a linear dependence of FM ordering on Fe content, $x$, above the percolation threshold, $x = x_{cr}$. The critical concentration of percolation threshold $x_{cr}^F \approx 0.09$ [26] for the case of face-centered cubic sub-lattices of magnetic ions. The percolation threshold is supposed to be essentially higher for AFM ordering, $x_{cr}^A \approx 0.48$ (see e.g. [27, 28] and refs therein). Note that superscripts "$F$" and "$A$" in $x_{cr}^{F,A}$ designate the critical concentrations related to FM and AFM ordering respectively. Thus, we assume that nonlinear magnetization expansion coefficient β critically depends on $x$. In particular, $\beta_L(x) = \beta_L(x - x_{cr}^A)/(1 - x_{cr}^A)$, $\beta_M(x) = \beta_M(x - x_{cr}^F)/(1 - x_{cr}^F)$ at content $x_{cr}^{F,A} \leq x \leq 1$; while $\beta_L(x) = 0$ and $\beta_M(x) = 0$ at $x < x_{cr}^{A,F}$.

In contrast to coefficient β, one can assume that the power expansion on $x$ is valid for biquadratic ME coefficients $\eta_{AFM}(x, R)$ and $\eta_{FM}(x, R)$, but they also tend to zero at $x \leq x_{cr}^F$, since the solid solution becomes nonmagnetic at $x \leq x_{cr}^F$. Hereinafter we assume that biquadratic ME coupling coefficients of FM and AFM have different signs, and in the most cases $\eta_{AFM} > 0$ and $\eta_{FM} < 0$ [29]. The assumption about different signs of $\eta_{AFM}(x)$ and $\eta_{FM}(x)$ agrees with Smolenskii and Chupis [30], Katsufuji and Takagi [31], and Lee et al [32]. In particular Smolenskii and Chupis [30], Katsufuji and Takagi [31] stated that it is natural to consider that the dielectric constant is dominated by the pair correlation between the nearest spins, which phenomenologically leads to the ME term $\eta P^2(M^2 - L^2)$.

For the composition $x > x_{cr}^A$, the temperature of the solid solution transition from the PM into AFM-ordering state is renormalized by the biquadratic ME coupling:

$$T_{PFY-PZT}^{AFM}(x, R) = T_N^{PFY}(R)\frac{(x - x_{cr}^A)}{(1 - x_{cr}^A)} - \frac{\eta_{AFM}(x)}{\alpha_{LT}}P_S^2(x, R). \tag{4}$$

Measured Néel temperatures for conventional bulk Pb(Fe$_{1/2}$Ta$_{1/2}$)O$_3$ and Pb(Fe$_{0.5}$Nb$_{0.5}$)O$_3$ are (143-170) K and (133-180) K correspondingly. Thus the values $T_N^{PFY}$ are "bare", because they are in fact shifted by the biquadratic ME coupling term $\eta_{AFM}P_S^2/\alpha_{LT}$ to lower temperatures, since $\eta_{AFM} > 0$. For nanostructured material the temperature depends on the size $R$ as $T_N^{PFY}(R) = T_N^Y\left(1 - \frac{R_{LY}}{R}\right)$. Corresponding size-dependence of polarization will be discussed below.

For the composition $x > x_{cr}^F$, the temperature of the solid solution transition from the PM into FM-ordering state is:

$$T_{PFY-PZT}^{FM}(x, R) = \theta_C^{PFY}(R)\frac{(x - x_{cr}^F)}{(1 - x_{cr}^F)} - \frac{\eta_{FM}(x, R)}{\alpha_{MT}}P_S^2(x, R). \tag{5}$$

Since PFT and PFN are antiferromagnets with negative temperature $\theta_C^{PFY}$ there should be strong enough biquadratic ME coupling term, $\eta_{FM}P_S^2/\alpha_{MT}$, that can strongly increases the FM-temperature for the solid solution up to the room and higher temperatures, since $\eta_{FM} < 0$. For nanostructured material the temperature $\theta_C^{PFY}$ is $R$-dependent as $\theta_C^{PFY}(R) = \theta_C^Y\left(1 - \frac{R_{MY}}{R}\right)$.

In Equations (4)-(5) the spontaneous polarization squire is

$$P_S^2(x, R) \approx \alpha_{PT}\left(T_{PFY-PZT}^{FE}(x, R) - T\right)/\beta_P(x), \tag{6}$$

where $T^{FE}_{PFY-PZT}(x,R)$ is given by Eqs.(3) allowing for $R$-dependence in the nanostructured case, and $\beta_P(x)$ depends on the composition x as $\beta_P(x) = \beta_P(0)(1 + \beta_x x + \beta_{xx} x^2 + ...)$. The power dependence of $\beta_P(x)$ on $x$ is in agreement with the well-known experimental results for the Pb-based solid solutions (see e.g. ref. [20] for PbZrTiO$_3$).

The substitution of $P_S^2(x,R)$, $\eta_{AFM}(x,R)$ and $\eta_{FM}(x,R)$ into Eqs.(4) and (5) leads to the evident dependences of the temperatures on the composition $x$ at given average radius $R$.

### 3.2. The impact of the size effect on the origin of the ferromagnetism in the solid solution PZT-PFT and PZT-PFN

The principal question is about the impact of the average size $R$ of the nanoregion on the possible origin of the ferromagnetism in the studied solid solution with one nonmagnetic component (PZT) and antiferromagnetic one (PFT or PFN) with $\theta_C^{PFY} < 0$. To answer the question let us analyze Eq.(5). For the case $x \leq x_{cr}^F$ any ferromagnetism is absent, since $\beta_M(x) = 0$. For the case of x increase and $x > x_{cr}^F$, the first term $\theta_C^{PFY}(R)\frac{(x - x_{cr}^F)}{(1 - x_{cr}^F)}$ decreases the Curie temperature $T^{FM}_{PFY-PZT}(x,R)$ because $\theta_C^{PFY} < 0$, but the second term $\frac{\eta_{FM}(x,R)}{\alpha_{MT}} P_S^2(x,R)$ can increase it, since $\eta_{FM}(x,R) < 0$ and the spontaneous polarization of PZT is 0.5 C/m$^2$ and it less than 0.10 C/m$^2$ for pure PFN at room temperature. PFT is in PE phase at room. Since the electro, magnetostriction, piezoelectric and piezomagnetic tensors strongly depend on the composition $x$, the ME coupling coefficients $\eta_{FM}$ and $\eta_{AFM}$ strongly vary with the composition x and size $R$ in agreement with Eq.(2b) and (2c). Moreover it is maximal for intermediate concentration x, where electrostriction, magnetostriction, piezoelectric coefficients are high and piezomagnetic coefficients are nonzero.

In result of the $P_S^2$ and $\eta_{FM}$ x-dependences, the term $\eta_{FM} P_S^2 / \alpha_{MT}$ can be 10 times higher for PFNx-PZT(1-x) and PFTx-PZT(1-x) than that for PFN or PFT, and for negative $\eta_{FM}$ it makes the sum $\left(\theta_C^{PFY}(R)\frac{(x - x_{cr}^F)}{(1 - x_{cr}^F)} - \frac{\eta_{FM}(x,R)}{\alpha_{MT}} P_S^2(x,R)\right)$ positive for intermediate x-compositions $x_{cr}^F \leq x < x_{cr}^A$. Immediately it leads to the positive FM Curie temperature in accordance with Eq.(5). So, negative $\eta_{FM}$ and high enough ratio $|\eta_{FM} P_S^2 / \alpha_{MT}|$ can give rise to the FM phase at intermediate x-compositions $x_{cr}^F \leq x < x_{cr}^{AFM}$ and lead to its disappearance with x increase at fixed $R$ value. It is worth to underline, that for the case $x=1$ it is possible to consider the limit $R \to \infty$, when there is only the first negative term in Eq.(5), so that FM phase is absent for PFT and PFN in agreement with experiment.

For arbitrary x value the phase diagrams of solid solutions PFNx-PZT(1-x) and PFTx-PZT(1-x) are shown in coordinates composition x – temperature $T$ in the **Figure 1a** and **1b** correspondingly.

One can see that the PE-FE boundary is well-fitted by the linear dependence, AFM-FE boundary - by the quasi-linear curve. Available experimental points for the FM-FE phase are not the boundary, but just indicate the region where it exists. So the FM boundary, that have a mountain-like form can reach higher temperatures and compositions x. Appeared that the temperature "height" of the FM-FE boundary is defined by the values of the negative ratio $\gamma_F = \eta_{FM}\alpha_{PT}/(\alpha_{MT}\beta_P)$ and "virtual" temperature $\theta_C^{PFY}$. The higher is the value $|\gamma_F|$ the higher and wider is the x-composition region of FM boundary. The smaller is the absolute value of the negative $\theta_C^{PFY}$, the wider is the FM boundary. The fitting values of $T_{PFY}^N$ appeared quite realistic, while $\theta_C^{PFY}$ are negative and its value is high enough, that is also in agreement with general knowledge [33]. The dimensionless ratio $\gamma_A = \eta_{AFM}\alpha_{PT}/(\alpha_{MT}\beta_P)$ is positive and is about unity. The ratio $\gamma_F$ is negative and high enough (<-30). As it follows from the **Figure 1** the theory given by solid lines describes the experimental points pretty good.

The next important issue is to analyze the fitting parameters used in the **Figure 1** and to understand if they can relate to conventional or nanostructured material, and to what typical sizes $R$ they correspond in the latter case. Appeared that the fitting values of $\gamma_F$ requires the factor $\left(1 + \frac{R_{\mu 1}}{R} + \left(\frac{R_{\mu 2}}{R}\right)^2\right)$ to be at least 10 times or higher for the coefficient $\eta_{FM}$ to be the same order as the realistic value of $2\times 10^{-22}$ s m/(VA) measured experimentally for PFN [15]. The condition leads to the inequality on the average size $R < 0.1 R_{\mu 1,2}$, i.e. can be true for nanostructured solid solution, but not for the homogeneous bulk. Finally the inequalities $R_{\mu 1,2} \geq 50-100$ nm and $R \leq 5-10$ nm should be valid, at the same time $R_{PY}$, $R_{LY}$ and $R_{MY}$ should be smaller than 1 – 2 nm in order not to shift essentially the temperatures $T_{PFY}(R) = T_Y^C\left(1 - \frac{R_{PY}}{R}\right)$, $T_N^{PFY}(R) = T_Y^C\left(1 - \frac{R_{LY}}{R}\right)$, $\theta_C^{PFY}(R) = \theta_Y^C\left(1 - \frac{R_{MY}}{R}\right)$ and polarization $P_S^2(x,R)$ from their values for homogeneous bulk. The small values of $R_{PY}$, $R_{LY}$ and $R_{MY}$ are required to provide high enough values of $P_S^2(x,R)$. Since the numerical values of $R_{\mu 1,2}$ and $R_{PY}$, $R_{LY}$, $R_{MY}$ are defined by different parameters, e.g. $R_{\mu 1} \propto \mu\left(\frac{Q_{ij}B_{ij} + Z_{ij}A_{ij}}{Q_{ij}Z_{ij}}\right)$, $R_{\mu 2} \propto \mu\sqrt{\frac{A_{ij}B_{ij}}{Q_{ij}Z_{ij}}}$ and $R_{PY} \propto \mu\left(\frac{Q_{ij}}{\alpha_{TY}T_Y^C}\right)$, $R_{LY} \propto \mu\left(\frac{\widetilde{Z}_{ij}}{\alpha_{LT}T_N^{PFY}}\right)$, $R_{MY} \propto \mu\left(\frac{Z_{ij}}{\alpha_{MT}\theta_Y^C}\right)$, to have respectively "big" value ($R_{\mu 1,2} \geq 50-100$ nm) and "small" ones ($R_{P,L,MY} < 1-2$ nm) are quite possible. At that the "optimal" nanoregion size should vary in the range $5R_Y < R < 0.1R_{\mu1,2}$. For example the sizes $R = 20-10$ nm, $R_{\mu2} \propto R_{\mu1} \approx 50-100$ nm, $R_{PY}$, $R_{LY}$, $R_{MY}$ less than 2 nm satisfy the all the conditions.

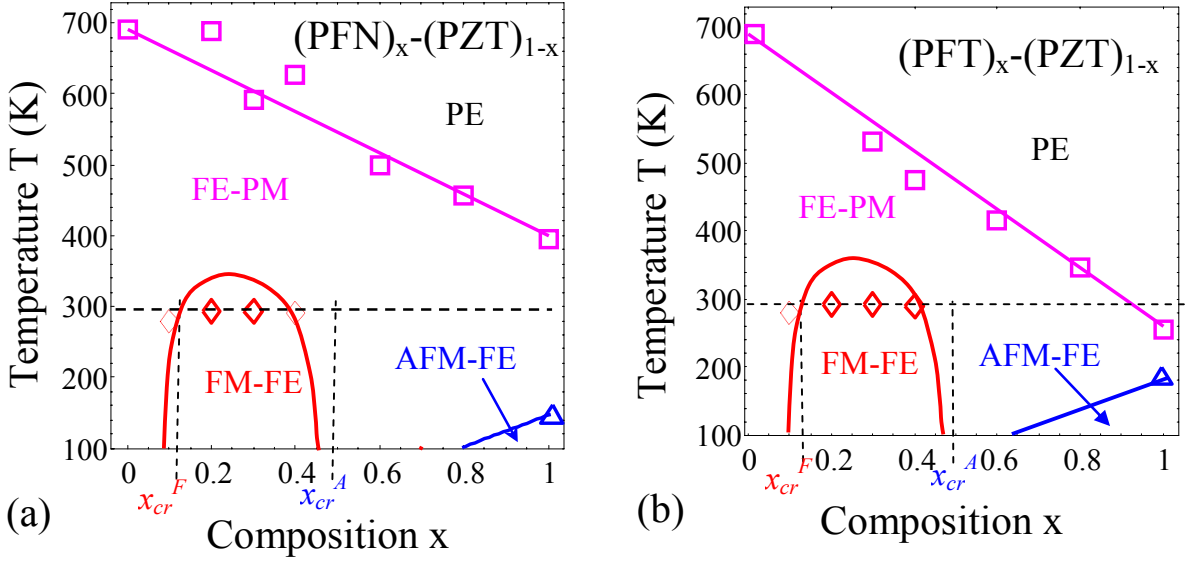

**Figure 1.** Phase diagrams in coordinates composition – temperature for solid solutions **(a)** Pb(Fe$_{1/2}$Nb$_{1/2}$)$_x$(Zr$_{0.53}$Ti$_{0.47}$)$_{1-x}$O$_3$ and **(b)** Pb(Fe$_{1/2}$Ta$_{1/2}$)$_x$(Zr$_{0.53}$Ti$_{0.47}$)$_{1-x}$O$_3$. Different symbols are experimental data collected from refs.[6, 8, 9, 20]. Boxes are data for PE-FE phase transition, triangles are AFM-PM boundary, diamonds show the points, where FM behaviour is observed. Solid curves are theoretical modelling for the percolation threshold concentrations $x_{cr}^F \approx 0.09$, $x_{cr}^A = 0.48$, $T_{PZT}^C = 690$ K, $P_S(0) = 0.5$ C/m$^2$ corresponds to PZT. Other parameters: **(a)** $T_N^{PFN} = 140$ K, $\gamma_A = 1.5$, $\gamma_F = -46$, $\theta_C^{PFN} = -550$ K **(b)** $T_N^{PFT} = 180$ K, $\gamma_A = 1$, $\gamma_F = -44$, $\theta_C^{PFT} = -500$ K [34].

### 3.3. Absence of the ferromagnetism in conventional PFN, PFT and in the solid solution PFNx-PT(1-x)

As it was shown experimentally ferromagnetism is absent in conventional PFN, PFT without any nanostructural elements as well as in the conventional solid solution PFNx-PT(1-x) [34], in contrast to nanostructured PFNx-PZT(1-x).

Firstly let us make estimations based on Eq.(5) to show that effective Curie temperature $T_{PFY}^{FM} = \theta_C^{PFY} - \frac{\eta_{FM}}{\alpha_{MT}} P_S^2$ is negative. Using the parameters $\theta_C^{PFY} = -(450-500)$ K (from experiment [34] and our fitting in the **Figure 1**), $P_S \approx (0.1 - 0.2)$ C/m$^2$, $\eta_{FM} \approx 2 \times 10^{-22}$ s m/(VA) [35] and $(\alpha_{MT})^{-1} \propto C_M \approx (0.2 - 0.25)$ K [15] we estimate the value $C_M \eta_{FM} P_S^2$ as $(40 - 250)$K, so that the temperature $T_{PFY}^{FM} < -200$ K. So, the ferromagnetism is absent in pure PFN and PFT.

The possible reason of the ferromagnetism absence in the solid solution PFNx-PT(1-x) is that PbTiO$_3$ (PT) has essentially smaller piezoelectric coefficients and dielectric permittivity at room temperature than PZT near the morphotropic boundary, meanwhile its spontaneous polarization can be even about 50% higher than that of PZT. Biquadratic ME coupling coefficient is absent for PT (x=0), but increases with x increase up to the value $\eta \approx 2\times 10^{-22}$ s m/(VA) for PFN. So, the biquadratic ME coupling coefficient is expected to be the same or even smaller than that for PFN. Using the parameters $\theta_C^{PFN} = -500$ K, "average" polarization $P_S \approx 0.5$ C/m$^2$, permittivity $\varepsilon=10^3$, "average" value $\tilde{\eta} \approx 10^{-22}$ s m/(VA) and $(\alpha_{MT})^{-1} \propto C_M \approx 0.25$ K, we estimated the $C_M \eta_{FM} P_S^2$ as -62.5 K making the temperature $T_{PFY}^{FM} < -390$ K. So, the ferromagnetism is absent in PFN$_{0.5}$–PT$_{0.5}$ in agreement with experiment [34].

Therefore these estimations lead to the conclusion that the biquadratic ME coefficients $\eta_{FM}(x)$ can be much smaller for PFNx-PT(1-x) than the ones for PFNx-PZT(1-x). Even higher polarization $P_S^2$ cannot make the term $\eta_{FM} P_S^2/\alpha_{MT}$ for PFNx-PT(1-x) as high as for PFTx-PZT(1-x). In result the term appeared not enough to make the sum positive and the conventional solution PFNx-PT(1-x) exhibit only AFM region below 100 K on the phase diagram [34].

## 4. Estimations of effective ME coupling coefficients

### 4.1. Estimation of linear ME coupling coefficients for conventional material at low temperatures

Linear ME coupling coefficients estimations are based on our previous results [23, 36], namely the bilinear coupling term is

$$\mu_{ij} \propto g_{jqs}^{(e)} d_{iqs}^{(m)} \qquad (7a)$$

Here $g_{ijk}^{(e)}$ and $d_{ijk}^{(m)}$ are coupling tensors of piezoelectric and piezomagnetic effects respectively.

In order to estimate the piezoeffects contribution to the bilinear coupling term of the solid solution, we should estimate the product, $g_{jqs}^{(e)} d_{iqs}^{(m)}$. In order to do this, we take into account that the corresponding strain is $u_{ij}^{(m)} \propto s_{ijps}^{-1} q_{pskl}^{(m)} M_k M_l \propto s_{ijkl} d_{qkl}^{(m)} M_q$; $s_{ijkl}$ are components of the elastic compliances tensor. Using of the values $s_{ijkl} \propto 5\times 10^{-12}$ Pa$^{-1}$, maximal piezoelectric coefficient corresponding to morphotropic boundary of Pb$_{0.5}$Zr$_{0.5}$TiO$_3$, $g_{ijk}^{(e)} \propto 0.1$ Vm/N at room temperature, high enough spontaneous magnetization, M~$5\times 10^3$ A/m (M~0.5emu/g~0.5A·m$^2$/kg, mass density 9.68×10$^3$ kg/m$^3$) and essential magnetostrictive strain value, $u^{(m)} \propto q^{(m)} M^2/s \sim 10^{-6}$, gives us $d^{(m)} \propto u^{(m)}/(sM) \sim 10^2$ Pa·m/A. Thus $g^{(e)} d^{(m)} \sim 10$ V/A and so $\mu_{ij} \propto 10$ V/A. To recalculate the value $\alpha$ in s/m, we can use that $\delta g_{ME} = \mu_{ij} M_i P_j = \alpha_{ij} H_i E_j$ and $P_i = \varepsilon_0 \varepsilon_{ij}^{(e)} E_j$, $M_i = \chi_{ij}^{(m)} H_j$ along with the values $\chi_{ij}^{(m)} < 10^{-3}$ and $\varepsilon_{ij}^{(e)} \approx 1.5\times 10^3$. Thus the linear ME coefficient value is $\alpha \propto \varepsilon_0 \varepsilon^{(e)} \chi^{(m)} \mu = 8.85\times 10^{-12} (F/m) \times 1.5\times 10^3 \times 10^{-3} \times 10 (V/A) \sim 1.3\times 10^{-10}$ s/m.

The value is at least 3 orders of magnitude smaller than the value of effective ME coefficient, $\alpha=1.3\times10^{-7}$ s/m, measured by Evans et al [8] for lamellar nanostructured material, where the ME coupling behavior appeared to be nonlinear. Actually, Evans et al [8] admitted that their effective coupling coefficient does not adhere to the strict definition of the linear ME coupling. On the other hand for single-phase multiferroics, linear ME coupling coefficients are typically of the order of $10^{-10}$ s/m, while for heterostructures values increase to of the order of $10^{-6}$ to $10^{-8}$ s/m [8]. Note that one order smaller piezoelectric coefficient $g_{ijk}^{(e)} \propto 10^{-3}$ Vm/N will lead the "standard" ME coefficient $\alpha \sim 10^{-11}$ s/m in Eq.(7a).

### 4.2. Estimation of effective ME coupling coefficients for nanostructured material

Let us estimate the effective ME coefficient for lamellar nanostructured material allowing for it possible nonlinearity and size effects. For the case of ferroelectric domain wall moving in external magnetic field H Evans et al calculated the ME coefficient from the formulae, $\alpha_{ij}^{eff} = \varepsilon E_i^{coer}/H_j^{crit}$, where the critical magnetic field $H_j^{crit} = 3$ kOe, coercive electric field and dielectric permittivity were taken from Sanchez et al [6] for x=0.4 as "bulk" coercive field $E_i^{coer} = 15$ kV/cm and bulk permittivity $\varepsilon=1200$. For the case of ferroelectric domain wall moving in external magnetic field $H$ more complex expression should be used for $\alpha_{ij}^{eff}$:

$$\alpha_{ij}^{eff}(R) \propto \alpha_{ij} + \beta_{ijk}(R)\varepsilon_{kl}^{eff} E_l^{coer} + \gamma_{ijk}(R)\chi_{kl}H_l^{crit} + \eta_{ijkl}(R)\varepsilon_{kn}^{eff} E_n^{coer}\chi_{lm}H_m^{crit} \qquad (7b)$$

Here $\varepsilon_{kl}^{eff}$ is the local dielectric permittivity, $\chi_{kl}$ is the magnetic permittivity, the coefficients $\beta_{ijk}(R) \propto \left(1 + \frac{R_{\mu\beta}}{R}\right)$, $\gamma_{ijk}(R) \propto \left(1 + \frac{R_{\mu\gamma}}{R}\right)$ and $\eta_{ijkl}(R) \propto \left(1 + \frac{R_{\mu 1}}{R} + \left(\frac{R_{\mu 2}}{R}\right)^2\right)$. For the case of the nanowire $R_{\mu\beta} = 2\mu \frac{s_{12}}{s_{11}}\frac{A_{33}}{Q_{11}}$, $R_{\mu\gamma} = 2\mu \frac{s_{12}}{s_{11}}\frac{B_{33}}{Z_{11}}$, $R_{\mu 1} = 2\mu \frac{s_{12}}{s_{11}}\left(\frac{Q_{11}B_{33} + Z_{11}A_{33}}{Q_{11}Z_{11}}\right)$ and $R_{\mu 2} = 2\mu \frac{s_{12}}{s_{11}}\sqrt{\frac{A_{33}B_{33}}{Q_{11}Z_{11}}}$. Note that the dependence on size R can increase the ratio $\alpha_{ij}^{eff}/\alpha_{ij}$ up to 10 or even $10^3$ times (more realistically in 10 – 100 times). Also one should take into account that the local dielectric permittivity $\varepsilon^{eff}$ in the immediate vicinity of the ferroelectric domain wall can be much higher that the bulk value $\varepsilon$. Indeed in accordance with the thermodynamic theory the ratio $\varepsilon_{ij}^{eff}/\varepsilon_{ij}$ diverges at the wall plane, but in reality it is finite (due to the presence of internal electric field), but can reach rather high values. So we see real possibilities to increase the ratio $\alpha_{ij}^{eff}/\alpha_{ij}$ up to 2-4 orders of magnitude due to the size and local permittivity increase. Thus the high value $\alpha^{eff} \sim (10^{-8} - 10^{-7})$s/m is reachable for nanostructured material in contrast to the conventional bulk value $\alpha \sim (10^{-11} - 10^{-10})$ s/m. Note, that since the coefficient $\eta_{ijkl}^{FM}(R)$ of nonlinear ME coupling has the same strong dependence on $R$ (see e.g. Eq.(2b)) it is not excluded that Evans et al [8] indeed observed quadratic ME effect.

## 5. Ferroelectric domains switching by a magnetic field

The estimation of $\alpha_{ij}$ is required to explain the ferroelectric domain structure switching by applied magnetic field of 3 kOe reported by Evans et al [8]. The nature of the phenomena is the following. Due to the presence of the strong bilinear ME coupling the magnetic field induces the electric field. For the magnetically isotropic media corresponding "acting" electric field is

$$E_i^{ME} = \mu\, M_i \equiv \frac{\alpha \chi^{(m)} H_i}{\varepsilon_0 \varepsilon^{(e)} \chi^{(m)}} = \frac{\alpha H_i}{\varepsilon_0 \varepsilon^{(e)}}, \qquad (8a)$$

where $M$ is magnetization, $H$ is magnetic field, $M_i = \chi^{(m)} H_i$ and the relation between $\mu$ and $\alpha$ magnitude is $\alpha = \varepsilon_0 \varepsilon^{(e)} \chi^{(m)} \mu$. Namely, variation of the thermodynamic potential (1) and (2a) via polarization leads to the equation of state:

$$\alpha_P P_i + \beta_{Pij} P_i P_j^2 + q_{mnil}^{(e)} u_{mn} P_l = -E_i^{ME} \qquad (8b)$$

Note, that Eq. (8b) defines the dependence of $E_i^{ME}$ and so $\alpha_P$ on x, R and T via the dependence of polarization on these quantities. If the component of the $E_i^{ME}$ conjugated with the spontaneous polarization component $P_i$ is higher than the critical field ($E_i^{cr}$) required for the domain wall motion, the polarization $P_i$ can be reversed by the field of appropriate direction. So that applied magnetic field can act as the source of ferroelectric domain structure triggering observed by Evans et al.

Note that the field $E_i^{cr}$ is typically much smaller than the thermodynamic coercive field. For PZT the critical field was measured as $E_i^{cr} = 120$ kV/cm [37], but it appeared to be much smaller for PFNx-PZT(1-x) or PFTx-PZT(1-x) due to the composition disorder, namely we will use the value $E_i^{cr} = 15$ kV/cm measured by Sanchez et al [6]. Using the estimation for $\alpha_{ij}^{eff} \approx 10^{-7}$ s/m and $\varepsilon$=1200 in Eq.(8a), one can lead to the conclusion that the critical magnetic field $H_i \approx 1.59 \times 10^5$ A/m is requited. The value is in a reasonable agreement with the fields of 3 kOe (that is equal to 2.3 $10^5$ A/m) applied by Evans et al [8].

**Figure 2** illustrates the dependence of the electric field $E^{ME}$ induced by the magnetic field H on bilinear ME coupling coefficient $\alpha^{eff}$ and H calculated from Eq.(8a). The contours of constant $E^{ME}$ has hyperbolic shape in coordinates ME coupling coefficient $\alpha$ - applied magnetic field $H$. The higher is $\mu$ the smaller field $H$ can induce the critical electric field that in turn can move the ferroelectric domain walls and switch their polarization. The contour $E^{ME} = E_{cr}$ was calculated for the critical field of 15 kV/cm. For $H$ and $\mu$ values below the contour the induced electric field $E^{ME} < E_{cr}$ and thus cannot change the domain structure. For $H$ and $\alpha^{eff}$ values above the contour the induced electric field $E^{ME} > E_{cr}$ and thus can change the domain structure.

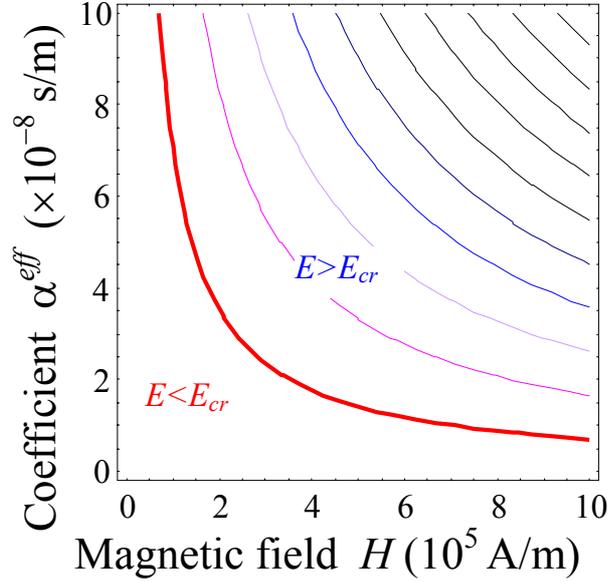

**Figure 2.** Contour map of the "acting" electric field is $E_i^{ME}$ in coordinates effective ME coupling coefficient $\alpha^{eff}$ - applied magnetic field $H$. The critical electric field is taken as 15 kV/cm.

## 6. Discussion and conclusion

The key point of the consideration performed in this paper was calculation of FM phase appearance in the nanostructured solid solution of $(PFY)_x(PZT)_{1-x}$ ($Y$ = Nb, Ta) at room temperature. Multiferroics PFN and PFT are the examples of ferroelectric antiferromagnets with $T_C \approx 380$ K, $T_N = 140$ K and $T_C \approx 250$ K, $T_N = 180$ K respectively and no FM transition was observed.

Room temperature FM and magnetoelectric phase with strong enough linear and squared ME coupling was revealed recently in the mentioned materials solid solution with PZT. The way, which permitted us to find out the mechanism of this wonderful phenomenon, was the following. Allowing for we were looking for FM phase at $T$ much higher than $T_N$, i.e. in paramagnetic phase, where susceptibility is described as $C/(T - \theta_C)$ ($\theta_C < 0$), it can be supposed that $\theta_C$ could be considered as some seeding temperature for FM phase, that has to be renormalized to positive value by some special mechanism. Since PZT is known to have strong piezoelectric and ferroelectric properties, which can increase linear and nonlinear ME coupling, we supposed that biquadratic ME coupling strongly enhanced by size effect could be the mechanism we were looking for. However, in pure PFN and PFT as well as in conventional $(PFN)_x(PT)_{1-x}$ without any nanostructure the addition given by this mechanism appeared to be not enough to make $T_C^{FM} > 0$ and so FM phase did not appear. In the case of nanostructured solid solutions with PZT high piezo- and electrostriction coefficients, $\eta_{FM}$ value along with large enough polarization it appeared possible to obtain FM phase even at room temperature. Note that there was discussion in literature (see e.g. [38, 39]) about existence of FM phase at low temperatures in different structures like $A^{2+}Fe_{1/2}B^{5+}_{1/2}O_3$ ($A^{2+}$ = Ba, Sr, Ca, Pb; $B^{5+}$ = Nb, Ta, W). To our mind the measurements of field-cooled susceptibility $1/\chi_{FC}$ at high temperature region as it was done in [34] for PFN will give the sign and value of temperature that correspond to $1/\chi_{FC} = 0$. Negative sign speaks in favours of antiferromagnet, while positive sign – of ferromagnetic material. The obtained value the $\theta_C \approx -(450-500)$ for PFN [34] is the direct evidence of FM phase absence in this material at any temperature $T \geq 0$ K.

Our consideration performed for nanostructured PFTx-PZT(1-x) and PFNx-PZT(1-x) made it possible to explain main experimental results observed by Evans et al [8] and Sanchez et al [6]. Namely, the anomalously large effective ME coupling coefficient $\alpha_{ij}^{eff}$ was shown to originate from size and local effects, which increase ME-coupling on 2-4 orders in comparison with conventional ceramics. The effective ME-coupling appeared to be close to quadratic ME effect, rather then linear one. The calculated values of $\alpha_{ij}^{eff}$ and critical magnetic fields necessary for ferroelectric domain walls motion was shown to be in reasonable agreement with experimentally observed quantities.

The quantitative estimations of linear ME coupling $\alpha$ via piezocoefficients for conventional ceramics (reported by Evans et al.[8]) had shown, that for PFTx-PZT(1-x) solid solution at $x = 0.3$ and $0.4$ $\alpha$ is about $1.3 \cdot 10^{-10}$ s/m, that is close to standard values measured by different authors [6,9]. The developed theory explains the absence of ferromagnetic phase in PFN, PFT and in the solid solution with PFN-PT [6, 9, 34]. We proposed the theoretical explanation of the solid solutions PFTx-PZT(1-x) and PFNx-PZT(1-x) phase diagrams dependence on the composition x with a special attention to the ferromagnetism and multiferroic properties for intermediate concentration x at room temperature.

## Acknowledgments

Authors acknowledge the support via bilateral SFFR-NSF project (US National Science Foundation under NSF-DMR-1210588 and State Fund of Fundamental State Fund of Fundamental Research of Ukraine, grant UU48/002). Authors are very grateful to Dr. Roman Kuzian for multiple discussions.